\documentclass{article}
\usepackage{amssymb}
\usepackage{amsxtra}
\usepackage{times}
\newtheorem{thm}{Theorem}[section]
\newtheorem{prop}[thm]{Proposition}
\newtheorem{lem}[thm]{Lemma}

\newtheorem{rem}[thm]{Remark}
\newcommand{\qed}{\hfill\hbox{\rule[-2pt]{3pt}{6pt}}}

\title{
\textbf{Generating Function Associated with the Determinant Formula}
\textbf{for the Solutions of the Painlev\'e II Equation}
}
\author{
{\normalsize Nalini JOSHI}\\
{\small School of Mathematics and Statistics F07, University of Sydney,}\\
{\small Sydney, NSW 2006, Australia}\\[5mm]
{\normalsize Kenji KAJIWARA}\\
{\small Graduate School of Mathematics, Kyushu University,}\\
{\small 6-10-1 Hakozaki, Higashi-ku, Fukuoka 812-8512, Japan}\\[5mm]
{\normalsize Marta MAZZOCCO}\\
{\small DPMMS, Wilberforce Road, Cambridge CB3 0WB, UK}\\
}

\date{}
\begin{document}
\maketitle
\begin{abstract}
In this paper we consider a Hankel determinant formula for generic
solutions of the Painlev\'e II equation.  We show that the generating
functions for the entries of the Hankel determinants are related to the
asymptotic solution at infinity of the linear problem of which the
Painlev\'e II equation describes the isomonodromic deformations.
\end{abstract}

\section{Introduction}
The Painlev\'e II equation (P$_{\rm II}$),
\begin{equation}
 \frac{d^2u}{dx^2}=2u^3-4xu+4\left(\alpha+\frac{1}{2}\right),\label{P2}
\end{equation}
where $\alpha$ is a parameter, is one of the most important equations in
the theory of nonlinear integrable systems.  It is well-known that
P$_{\rm II}$ admits unique rational solution when $\alpha$ is a
half-integer, and one-parameter family of solutions expressible in terms
of the solutions of the Airy equation when $\alpha$ is an
integer. Otherwise the solution is
non-classical \cite{Murata,Noumi-Okamoto, Watanabe-Umemura}.

The rational solutions for P$_{\rm II}$(\ref{P2}) are expressed as logarithmic
derivative of the ratio of certain special polynomials, which are called the
``{\sl Yablonski-Vorob'ev polynomials}'',  \cite{Vo,Ya}. Yablonski-Vorob'ev polynomials
admit two determinant formulas, namely, Jacobi-Trudi type and Hankel
type. The latter is described as follows: 
For each positive integer $N$, the 
unique rational solution for $\alpha = N+1/2$ is given by    
\begin{displaymath}
u = \dfrac{d}{dx} \log \dfrac{\sigma_{N+1}}{\sigma_N} ,
\end{displaymath}
where $\sigma_N$ is the Hankel determinant 
\begin{displaymath}
\sigma_N = \left|
\begin{array}{cccc}
a_0 & a_1 & \cdots & a_{N-1} \\
a_1 & a_2 & \cdots & a_N \\
\vdots & \vdots & \ddots & \vdots \\
a_{N-1} & a_N & \cdots & a_{2N-2} 
\end{array}
\right|  ,
\end{displaymath}
with $a_n = a_n(x)$ being polynomials defined by the 
recurrence relation 
\begin{equation} \label{eqn:recurrence_rational} 
\begin{array}{rcl} 
    a_0 &=& x, \qquad a_1 = 1,  \\
\vspace{-0.3cm} & &             \\ 
a_{n+1} &=& \dfrac{d a_n}{dx} + \displaystyle 
\sum_{k=0}^{n-1} a_k \, a_{n-1-k}.  
\end{array} 
\end{equation} 

The Jacobi-Trudi type formula implies that the Yablonski-Vorob'ev polynomials
are nothing but the specialization of the Schur
functions \cite{KO:p2}. Then, what does the Hankel determinant formula
mean? In order to answer this question, a generating function for $a_n$
is constructed in  \cite{IKN}:
\begin{thm} \cite{IKN} \label{thm:IKN} 
Let $\theta(x,t)$ be an entire function of two variables 
defined by 
\begin{equation}  \label{IKN:theta}
\theta(x,t) = \exp\left(2 t^3/3 \right)\, \mathrm{Ai}(t^2-x),
\end{equation}  
where $\mathrm{Ai}(z)$ is the Airy function. 
Then there exists an asymptotic expansion 
\begin{equation} \label{IKN:main}
\dfrac{\partial}{\partial t} \log \theta(x,t) \sim 
\sum_{n=0}^{\infty} a_n(x) \, (-2t)^{-n},
\end{equation} 
as $t \to \infty$ in any proper subsector of the sector 
$|\arg t|< \pi/2$.  
\end{thm}  
This result is quite suggestive, because it shows that the Airy 
functions enter 
twice in the theory of classical solutions of the P$_{\rm II}$: 
\begin{enumerate}
\item in the formula  \cite{Er}
$$ u = \dfrac{d}{dx} \log \mathrm{Ai}\left(2^{1/3} x \right), 
\qquad \alpha = 0.
$$
the one parameter family of classical solutions of 
P$_{\rm II}$  for integer values of $\alpha$ is expressed by Airy functions,
\item in formulae  (\ref{IKN:theta}), (\ref{IKN:main}) 
the Airy functions generate the entries 
of determinant formula for the rational solutions. 
\end{enumerate}

In this paper we clarify the nature of this phenomenon.  First, we
reformulate the Hankel determinant formula for generic, namely
non-classical, solutions of P$_{\rm II}$ already found in
 \cite{KM:p2_generic,KMNOY:det}. We next construct generating functions
for the entries of our Hankel determinant formula.  We then show that
the generating functions are related to the asymptotic solution at
infinity of the isomonodromic problem introduced by Jimbo and Miwa
\cite{JMU}. More explicitly,  the generating functions we construct are 
represented formally by series in powers of a variable $t$ that does not 
appear in the second Painlev\'e equation. We show that they satisfy two
Riccati equations, one in the $x$ variable of P$_{\rm II}$, the other in 
the auxiliary variable $t$. These Riccati equations simultaneously linearise
to the two linear systems whose compatibility is given by P$_{\rm II}$.
This is the first time in the literature, to our
knowledge, that the construction of the isomonodromic deformation problem 
has been carried out by starting directly from the Painlev\'e equation of 
interest.

This result explains the appearance of the Airy functions in Theorem
\ref{thm:IKN}.  In fact, for rational solutions of P$_{\rm II}$, the
asymptotic solution at infinity of the isomonodromic problem is indeed
constructed in terms of Airy functions  \cite{J1,J2,O1}.
%%% Comment by Kajiwara:
%%% In  \cite{Jimbo} only PVI, PV and PIII are considered but not PII.
%%% Okamoto have solved the isomonodoromy problem for the simplest
%%% rational solution of PII -- PIV but this is not published.
%%% Please confirm the case of PII. If you could not confirm, it might
%%% be btter to delete this paragraph.

We expect that the generic solutions of the so-called Painlev\'e II
hierarchy  \cite{AS,Ai,FN} should be expressed by some Hankel determinant
formula.  Of course the generating functions for the entries of Hankel
determinant should be related to the asymptotic solution at infinity of
the isomonodromic problem for the Painlev\'e II hierarchy.  We also
expect that the similar phenomena can be seen for other Painlev\'e
equations.  We shall discuss these generalizations in future
publications.

\vskip 0.2 cm 
\noindent{\bf Acknowledgements} The authors thank Prof. H. Sakai for
informing them of references  \cite{J1,J2}. They also thank
Prof. K. Okamoto for discussions and encouragement.  M.M. acknowledges
the support from the Engineering and Physical Sciences Research Council
Fellowship \#GR/S48424/01. K.K. acknowledges the support from the
scientist exchange program between Japan Society for the Promotion of
Science and Australian Academy of Science \#0301002 and the JSPS
Grant-in-Aid for Scientific Research (B) \#15340057.

\section{Hankel Determinant Formula and Isomonodromic Problem}
\subsection{Hankel Determinant Formula}
We first prepare the Hankel determinant formula for generic solutions
for P$_{\rm II}$ (\ref{P2}). To show the parameter dependence
explicitly, we denote equation (\ref{P2}) as P$_{\rm II}[\alpha]$.
The formula is based on the fact that the $\tau$ functions for P$_{\rm
II}$ satisfy the Toda equation,
\begin{equation}
 \sigma_n''\sigma_n - \left(\sigma_n'\right)^2 = \sigma_{n+1}\sigma_{n-1},\quad n\in\mathbb{Z}, \quad
{}'=d/dx.\label{Toda:sigma}
\end{equation}
Putting $\tau_n=\sigma_n/\sigma_0$ so that the $\tau$ function is
normalized as $\tau_0=1$, equation (\ref{Toda:sigma}) is rewritten as
\begin{equation}
 \tau_n''\tau_n-\left(\tau_n'\right)^2=\tau_{n+1}\tau_{n-1}-\varphi\psi\tau_n^2,\quad
\tau_{-1}=\psi,\quad\tau_0=1,\quad\tau_1=\varphi,\quad n\in\mathbb{Z}.\label{Toda:tau}
\end{equation}
Then it is known that $\tau_n$ can be written in terms of Hankel determinant as
follows \cite{KMNOY:det}:
\begin{prop}\label{prop:det}
 Let $\{a_n\}_{n\in \mathbb{N}}$,
$\{b_n\}_{n\in\mathbb{N}}$ be the sequences defined
 recursively as
\begin{equation}
 a_n=a_{n-1}^\prime + \psi\sum_{i+j=n-2\atop i,j\geq 0}a_ia_j,
\quad  b_n=b_{n-1}^\prime + \varphi\sum_{i+j=n-2\atop i,j\geq 0}b_ib_j,\quad a_0=\varphi,\quad b_0=\psi.
\label{Toda:rec}
\end{equation}
For any $N\in\mathbb{Z}$, we define Hankel determinant $\tau_N$ by
\begin{equation}
 \tau_N=\left\{
\begin{array}{ll}
\det(a_{i+j-2})_{i,j\leq N}&N>0,\\
1, &N=0,\\
\det(b_{i+j-2})_{i,j\leq |N|}&N<0.
\end{array}
\right.\label{Toda:det}
\end{equation}
Then $\tau_N$ satisfies equation (\ref{Toda:tau}).
\end{prop}
Since the above formula involves two arbitrary functions $\varphi$ and
$\psi$, it can be regarded as the determinant formula for general
solution of the Toda equation. Imposing appropriate conditions on
$\varphi$ and $\psi$, we obtain determinant formula for the solutions of
P$_{\rm II}$:
\begin{prop}
Let $\psi$ and $\varphi$ be functions in $x$ satisfying
\begin{eqnarray}
&& \frac{\psi''}{\psi} = \frac{\varphi''}{\varphi} = -2\psi\varphi + 2x,\label{cond1:p2_generic}\\
&& \varphi'\psi - \varphi\psi'=2\alpha, \label{cond2:p2_generic}
\end{eqnarray}
Then we have the following:
\begin{enumerate}
 \item $u_0=(\log\varphi)'$ satisfies P$_{\rm II}[\alpha]$.
 \item $u_{-1}=-(\log\psi)'$ satisfies P$_{\rm II}[\alpha-1]$.
 \item $u_N=\left(\log\frac{\tau_{N+1}}{\tau_N}\right)'$, where $\tau_N$
       is defined by equation (\ref{Toda:det}), satisfies  P$_{\rm II}[\alpha+N]$.
\end{enumerate}
\end{prop}
\noindent \textbf{Proof.} (i) and (ii) can be directly checked by using the
relations (\ref{cond1:p2_generic}) and (\ref{cond2:p2_generic}). Then
(iii) is the reformulation of Theorem 4.2 in  \cite{KM:p2_generic}. \qed
\subsection{Riccati Equations for Generating Functions}
Consider the generating functions for the entries as the following formal 
series
\begin{equation}\label{coefs}
 F_\infty(x,t)=\sum_{n=0}^\infty a_n(x)~t^{-n},\quad
 G_\infty(x,t)=\sum_{n=0}^\infty b_n(x)~t^{-n}.
\end{equation}
It follows from the recursion relations (\ref{Toda:rec})
that the generating functions formally satisfy the Riccati equations. In fact,
multiplying the recursion relations (\ref{Toda:rec}) by $t^{-n}$ and
take the summation from $n=0$ to $\infty$, we have:

\begin{prop}\label{prop:riccati:x}
The generating functions $F_\infty(x,t)$ and $G_\infty(x,t)$ formally 
satisfy the Riccati equations
\begin{eqnarray}
&&t\frac{\partial F}{\partial x}=-\psi F^2 +t^2 F - t^2 \varphi, 
\label{riccati:Fx}\\
&&t\frac{\partial G}{\partial x}=-\varphi G^2 +t^2 G - t^2 \psi,
 \label{riccati:Gx}
\end{eqnarray}
respectively.
\end{prop}
Since $F_\infty$ and $G_\infty$ are defined as the formal power series around
$t\sim\infty$, it is convenient to derive the differential equations
with respect to $t$. In order to do this, the following auxiliary
recursion relations are useful.
\begin{lem}\label{rec:aux}
 Under the condition (\ref{cond1:p2_generic}) and
 (\ref{cond2:p2_generic}), $a_n$ and $b_n$ ($n\geq 0$) satisfy the following
 recursion relations,
\begin{eqnarray}
&& \left(\psi a_{n+2}-\psi' a_{n+1}\right)'=2(n+1)\psi a_{n},\label{aux:a}\\
&& \left(\varphi b_{n+2}-\varphi' b_{n+1}\right)'=2(n+1)\varphi b_{n},\label{aux:b}
\end{eqnarray}
respectively.
\end{lem}
We omit the details of the proof of Lemma \ref{rec:aux}, because it is
proved by straight but tedious induction. Multiplying equations (\ref{aux:a})
and (\ref{aux:b}) by $t^{-n}$ and taking summation over $n=0$ to
$\infty$, we have the following differential equations for $F_\infty$ and $G_\infty$:
\begin{lem}\label{riccati:tx}
 The generating functions $F_\infty$ and $G_\infty$ formally satisfy 
the following differential equations,
\begin{eqnarray}
&&  2\psi t\frac{\partial F}{\partial t}=t(\psi'-t\psi)\frac{\partial F}{\partial x} +(\psi''t-\psi' t^2+2\psi)F
+t^2(\psi\varphi'+\psi'\varphi),\label{Ft_and_Fx}\\
&&  2\varphi t\frac{\partial G}{\partial t}=t(\varphi'-t\varphi)\frac{\partial G}{\partial x} 
+(\varphi''t-\varphi' t^2+2\varphi)G+t^2(\psi\varphi'+\psi'\varphi),\label{Gt_and_Gx}
\end{eqnarray}
respectively.
\end{lem}
Eliminating $x$-derivatives from equations (\ref{riccati:Fx}),
(\ref{Ft_and_Fx}), and equations (\ref{riccati:Gx}), (\ref{Gt_and_Gx}), respectively, we
obtain the Riccati equations with respect to $t$:
\begin{prop}\label{prop:riccati:t}
 The generating functions $F_\infty$ and $G_\infty$ formally satisfy the following Riccati
 equations,
\begin{eqnarray}
&& 2t\frac{\partial F}{\partial t}=-(\psi'-t\psi) F^2 + \left(\frac{\psi''}{\psi}t+2-t^3\right)F
+t^2(\varphi'+t\varphi) ,\label{riccati:Ft}\\
&& 2t\frac{\partial G}{\partial t}=-(\varphi'-t\varphi) G^2 + \left(\frac{\varphi''}{\varphi}t+2-t^3\right)G
+t^2(\psi'+t\psi) ,\label{riccati:Gt}
\end{eqnarray}
respectively.
\end{prop}
\subsection{Isomonodromic Problem}
The Riccati equations for $F_\infty$ equations (\ref{riccati:Fx}) and (\ref{riccati:Ft}) are
linearized by standard technique, which yield isomonodromic problem
for P$_{\rm II}$. It is easy to derive the following theorem from the
Proposition \ref{prop:riccati:x} and \ref{prop:riccati:t}:
\begin{thm}\label{thm:main}
\begin{enumerate}
 \item It is possible to introduce the functions $Y_1(x,t)$, $Y_2(x,t)$ consistently as
\begin{equation}
  F_\infty(x,t) = \frac{t}{\psi}\left(\frac{1}{Y_1}\frac{\partial Y_1}{\partial x}+\frac{t}{2}\right)
   = \frac{2t}{\psi'-t\psi}\left[\frac{1}{Y_1}\frac{\partial Y_1}{\partial t}+\frac{1}{4}\left(\frac{\psi''}{\psi}
-t^2\right)\right],\label{YxYt}
\end{equation}
\begin{equation}
 Y_2 = \frac{1}{\psi}\left(\frac{\partial Y_1}{\partial x}+\frac{tY_1}{2}\right).\label{Y2}
\end{equation}
Then $Y_1$ and $Y_2$ satisfy the following linear system for $Y=\left(\begin{array}{c}Y_1 \\ Y_2
	   \end{array}\right)$:
\begin{eqnarray}
&& \frac{\partial}{\partial t}Y = AY,\quad A=\left(
\begin{array}{cc}
\smallskip
{\displaystyle \frac{t^2}{4}-\frac{z}{2}-\frac{x}{2}}& {\displaystyle -\frac{\psi}{2}(t+u_{-1}) }\\
{\displaystyle \frac{1}{\psi}\left\{(u_{-1}-t)\frac{z}{2}+\alpha\right\}} & 
{\displaystyle -\frac{t^2}{4}+\frac{z}{2}+\frac{x}{2}}
\end{array}
\right) ,
\label{Lax:a:t}\\
&&\frac{\partial}{\partial x}Y = BY,\quad
B=\left(
\begin{array}{cc}
\smallskip
{\displaystyle  -\frac{t}{2}}& {\displaystyle \psi}\\
{\displaystyle \frac{z}{\psi} }&{\displaystyle  \frac{t}{2}}
\end{array}
\right), \label{Lax:a:x}
\end{eqnarray}
where $z=-\psi\varphi$. 
 \item Similarly, it is possible to introduce the functions $Z_1(x,t)$, $Z_2(x,t)$ consistently as
\begin{equation}
  G_\infty(x,t) = \frac{t}{\varphi}\left(\frac{1}{Z_1}\frac{\partial Z_1}{\partial x}+\frac{t}{2}\right)
   = \frac{2t}{\varphi'-t\varphi}\left[\frac{1}{Z_1}\frac{\partial Z_1}{\partial t}+\frac{1}{4}\left(\frac{\varphi''}{\varphi}-t^2\right)\right],\label{ZxZt}
\end{equation}
\begin{equation}
 Z_2 = \frac{1}{\varphi}\left(\frac{\partial Y_1}{\partial x}+\frac{tY_1}{2}\right).\label{Z2}
\end{equation}
Then $Z_1$ and $Z_2$ satisfy the following linear system for
$Z=\left(\begin{array}{c}Z_1 \\ Z_2 \end{array}\right)$:
\begin{eqnarray}
&& \frac{\partial}{\partial t}Z = CZ,\quad 
C=\left(
\begin{array}{cc}
\smallskip
{\displaystyle \frac{t^2}{4}-\frac{z}{2}-\frac{x}{2}}& {\displaystyle -\frac{\varphi}{2}(t-u_{0}) }\\
{\displaystyle -\frac{1}{\varphi}\left\{(u_{0}+t)\frac{z}{2}+\alpha\right\}} & 
{\displaystyle -\frac{t^2}{4}+\frac{z}{2}+\frac{x}{2}}
\end{array}
\right) ,\label{Lax:b:t}\\
&&\frac{\partial}{\partial x}Z = DY,\quad D=\left(
\begin{array}{cc}
\smallskip
{\displaystyle  -\frac{t}{2}}& {\displaystyle \varphi}\\
{\displaystyle \frac{z}{\varphi} }&{\displaystyle  \frac{t}{2}}
\end{array}
\right).\label{Lax:b:x}
\end{eqnarray}
\end{enumerate}
\end{thm}
\begin{rem}
The linear systems  (\ref{Lax:a:t}), (\ref{Lax:a:x}) and (\ref{Lax:b:t}), 
(\ref{Lax:b:x}) are the isomonodoromic problems for P$_{\rm II}[\alpha-1]$ 
and P$_{\rm II}[\alpha]$, respectively  \cite{JMU}. For example, compatibility
       condition for equations (\ref{Lax:a:t}) and (\ref{Lax:a:x}),
       namely, 
\begin{displaymath}
  \frac{\partial A}{\partial x} - \frac{\partial B}{\partial t} + [A,B]=0,
\end{displaymath}
gives
\begin{equation}
 \left\{
\begin{array}{l}
\smallskip
{\displaystyle \frac{dz}{dx}=-2u_{-1}z-2\alpha},\\
\smallskip
{\displaystyle \frac{du_{-1}}{dx}=u_{-1}^2-2z-2x},\\
{\displaystyle u_{-1}=-\frac{1}{\psi}\frac{d\psi}{dx}},
\end{array}
\right. 
\end{equation}
which yields P$_{\rm II}[\alpha-1]$ for $u_{-1}$. This fact also
       guarantees the consistency of two expressions for $F_\infty$ in
       terms of $Y_1$ in equation (\ref{YxYt}). Similar remark holds true for
       $G_\infty$ and $Z_1$. 
\end{rem}
\begin{rem}\label{rem:FandY}
$F_\infty$ and $G_\infty$ are also expressed as,
\begin{equation}
 F_\infty=t~\frac{Y_2}{Y_1},\quad G_\infty=t~\frac{Z_2}{Z_1},
\end{equation}
respectively. Conversely, it is obvious that for any solution $Y_1$ and $Y_2$ for the
       linear system (\ref{Lax:a:t}) and (\ref{Lax:a:x}), $F=tY_2/Y_1$
       satisfies the Riccati equations (\ref{riccati:Fx}) and
       (\ref{riccati:Ft}) (Similar for $G$). 
\end{rem}
\begin{rem}
Theorem \ref{thm:IKN} is recovered by putting $\psi=1$, $\varphi=x$.
\end{rem}
\begin{rem}
$Y_1$ can be formally expressed in terms of $a_n$ by 
using equation (\ref{YxYt}) as
\begin{equation}
 Y_1={\rm const.}\times 
\exp\left(\frac{1}{12}t^3-\frac{x}{2}t\right) ~t^{-\alpha}~
\exp\left[\frac{1}{2}\sum_{n=1}^{\infty} 
\frac{\psi a_{n+1}-\psi' a_n}{n}~t^{-n}\right].\label{Y_and_an}
\end{equation}
which coincides with known asymptotic behavior of $Y_1$ around 
$t\sim\infty$ \cite{JMU}.
\end{rem}

\section{Solutions of Isomonodromic Problems and Determinant Formula}
In the previous section we have shown that the generating functions 
$F_\infty$ and $G_\infty$ formally satisfy the Riccati equations
(\ref{riccati:Fx},\ref{riccati:Ft}) and (\ref{riccati:Gx},\ref{riccati:Gt}), 
and that their linearization yield isomonodromic problems (\ref{Lax:a:t},
\ref{Lax:a:x}) and (\ref{Lax:b:t},\ref{Lax:b:x}) for P$_{\rm II}$. 
Now let us consider the
converse. We start from the linear system (\ref{Lax:a:t}) and
(\ref{Lax:a:x}). We have two linearly independent solutions around
$t\sim\infty$, one of which is related with $F_\infty(x,t)$. Then, what is
another solution? In fact, it is well-known that linear system
(\ref{Lax:a:t}) and (\ref{Lax:a:x}) admit the formal solutions around
$t\sim\infty$ of the form \cite{JMU},
 \begin{eqnarray}
 Y^{(1)}&=&\left(\begin{array}{c}Y^{(1)}_1 \\Y^{(1)}_2 \end{array}\right)
=\exp\left(\frac{t^3}{12}-\frac{xt}{2}\right)t^{-\alpha}\left[\left(\begin{array}{c}1 \\0 \end{array}\right)
+\left(\begin{array}{c}y^{(1)}_{11} \\y^{(1)}_{21} \end{array}\right)t^{-1}
+\left(\begin{array}{c}y^{(1)}_{12} \\y^{(1)}_{22} \end{array}\right)t^{-2}+\cdots\right],\\
 Y^{(2)}&=&\left(\begin{array}{c}Y^{(2)}_1 \\Y^{(2)}_2 \end{array}\right)
=\exp\left(-\frac{t^3}{12}+\frac{xt}{2}\right)t^{\alpha}\left[\left(\begin{array}{c}0 \\1 \end{array}\right)
+\left(\begin{array}{c}y^{(2)}_{11} \\y^{(2)}_{21} \end{array}\right)t^{-1}
+\left(\begin{array}{c}y^{(2)}_{12} \\y^{(2)}_{22} \end{array}\right)t^{-2}+\cdots\right].
\end{eqnarray}
From Remark \ref{rem:FandY} we see that there are two possible
power-series solutions for the Riccati equation (\ref{riccati:Ft}) of
the form,
\begin{eqnarray}
&& Y^{(1)}\rightarrow F^{(1)}=t~\frac{Y_2^{(1)}}{Y_1^{(1)}}
=t~\frac{y_{21}^{(1)}t^{-1}+\cdots}{1+y_{11}^{(1)}t^{-1}+\cdots}
= c_0 + c_1t^{-1}+\cdots,\\
&&Y^{(2)}\rightarrow F^{(2)}=t~\frac{Y_2^{(2)}}{Y_1^{(2)}}
=t~\frac{1+y_{21}^{(2)}t^{-1}+\cdots}{y_{11}^{(2)}t^{-1}+\cdots}
= t^2(d_0 + d_1t^{-1}+\cdots).
\end{eqnarray}
The above two possibilities of power-series solutions for the Riccati equations
are verified directly as follows:
\begin{prop}
 The Riccati equation (\ref{riccati:Ft}) admits only the following two kinds of power-series 
 solutions around $t\sim\infty$:
\begin{equation}
 F^{(1)} = \sum_{n=0}^{\infty} c_n~t^{-n},\quad
 F^{(2)} = t^2\sum_{n=0}^{\infty} d_n~t^{-n}.\label{series:F}
\end{equation}
\end{prop}
\noindent\textbf{Proof.} We substitute the expression,
\begin{equation}
 F=t^{\rho}\sum_{n=0}^\infty c_n~t^{-n},
\end{equation}
for some integer $\rho$ to be determined, into the Riccati equation (\ref{riccati:Ft}). We have:
\begin{eqnarray*}
 \sum_{n=0}^{\infty}2(\rho-n)c_nt^{\rho+1-n}
&=& \sum_{n=0}^{\infty}\psi'\left( \sum_{k=0}^{n}c_kc_{n-k}\right)t^{2\rho-2n}
- \sum_{n=0}^{\infty}\psi\left( \sum_{k=0}^{n}c_kc_{n-k}\right)t^{2\rho+1-2n}\\
&&+ \sum_{n=0}^{\infty}\left(\frac{\psi''}{\psi}+2\right)c_nt^{\rho-n}
- \sum_{n=0}^{\infty}c_nt^{\rho+3-n}+t^2(\varphi'+t\varphi)
\end{eqnarray*}
The leading order should be one of $t^{2\rho+1}$, $t^{\rho+3}$ and
$t^3$. Investigating the balance of these terms, we have 
$\rho=0$ or $\rho=2$.\qed\\[3mm]
We also have the similar result for the solution of the Riccati equation (\ref{riccati:Gt}):
\begin{prop}
 The Riccati equation (\ref{riccati:Gt}) admits only the following two kinds of power-series 
 solutions around $t\sim\infty$:
\begin{equation}
 G^{(1)} = \sum_{n=0}^{\infty} e_n~t^{-n},\quad
 G^{(2)} = t^2\sum_{n=0}^{\infty} f_n~t^{-n}.\label{series:G}
\end{equation}
\end{prop}

It is obvious that $F^{(1)}$ and $G^{(1)}$ are nothing but our
generating functions $F_\infty$ and $G_\infty$, respectively. Then, what are $F^{(2)}$
and $G^{(2)}$? In the following, we present two observations regarding
this point.  First, there are unexpectedly simple relations among those functions:
\begin{prop}\label{prop:FandG}
The following relations holds.
\begin{equation}
F^{(2)}(x,t)=\frac{t^2}{G^{(1)}(x,-t)} ,\quad G^{(2)}(x,t)=\frac{t^2}{F^{(1)}(x,-t)}.
\end{equation}
\end{prop}
\noindent{\bf Proof.} Substitute $F(x,t)=\frac{t^2}{g(x,t)}$ into
equation (\ref{riccati:Ft}). This gives equation (\ref{riccati:Gt}) for
$G(x,t)=g(x,-t)$ by using the relation (\ref{cond1:p2_generic}). Choosing
$g(x,t)=G^{(1)}(x,t)$, $F(x,t)$ must be $F^{(2)}(x,t)$, since its leading
order is $t^2$. We obtain the second equation by the similar argument.\qed
\\[3mm]

Second, $F^{(2)}(x,t)$ and $G^{(2)}(x,t)$ are also interpreted as
generating functions of entries of Hankel determinant formula for
P$_{\rm II}$. Recall that the determinant formula in Proposition
\ref{prop:det} is for the $\tau$ sequence $\tau_n=\sigma_n/\sigma_0$ so
that it is normalized as $\tau_0=1$. We show that $F^{(2)}(x,t)$ and 
$G^{(2)}(x,t)$ correspond
to different normalizations of $\tau$ sequence:
\begin{prop}\label{prop:F2andtau}
Let 
\begin{eqnarray}
&& F^{(2)}(x,t)=-\frac{t^2}{\psi^2}\sum_{n=0}^\infty d_n~(-t)^{-n},\label{F2:dn}\\
&& G^{(2)}(x,t)=-\frac{t^2}{\varphi^2}\sum_{n=0}^\infty f_n~(-t)^{-n},\label{G2:fn}
\end{eqnarray}
be formal solutions of the Riccati equations
(\ref{riccati:Fx}),(\ref{riccati:Ft}) and (\ref{riccati:Gx}),
(\ref{riccati:Gt}), respectively.  We put
\begin{eqnarray}
&&\kappa_{-n-1}=\det(d_{i+j})_{i,j=1,\ldots,n}\quad (n>0),\quad \kappa_{-1}=1,\label{kappaandtau}\\
&&\theta_{n+1}=\det(f_{i+j})_{i,j=1,\ldots,n}\quad (n>0),\quad \theta_1=1.\label{thetaandtau}
\end{eqnarray}
Then $\kappa_n$ and $\theta_n$ are related to $\tau_n$ as
\begin{eqnarray}
&& \kappa_n=\frac{\tau_{n}}{\psi}=\frac{\tau_{n}}{\tau_{-1}}\quad (n<0),\\
&& \theta_n=\frac{\tau_{n}}{\varphi}=\frac{\tau_{n}}{\tau_{1}}\quad (n>0).
\end{eqnarray}
\end{prop}
To prove Proposition \ref{prop:F2andtau}, we first derive recurrence
relations that characterize $d_n$ and $f_n$. By substituting
equations (\ref{F2:dn}) and (\ref{G2:fn}) into the Riccati equations
(\ref{riccati:Fx}) and (\ref{riccati:Gx}), respectively, we easily
obtain the following lemma:
\begin{lem}\label{lem:recdn}
\begin{enumerate}
 \item $d_0$ and $d_1$ are given by $d_0=-\psi$ and $d_1=\psi'$,
       respectively. For $n\geq 2$, $d_n$ are characterized by the
       recursion relation,
\begin{equation}
 d_n=d_{n-1}'+\frac{1}{\psi}\sum_{k=2}^{n-2}d_kd_{n-k},\quad
d_2=\frac{\psi''\psi-(\psi')^2+\varphi\psi^3}{\psi}.
\end{equation}
 \item $f_0$ and $f_1$ are given by $d_0=-\varphi$ and $d_1=\varphi'$,
       respectively. For $n\geq 2$, $f_n$ are characterized by the
       recursion relation,
\begin{equation}
 f_n=f_{n-1}'+\frac{1}{\varphi}\sum_{k=2}^{n-2}f_kf_{n-k},\quad
f_2=\frac{\varphi''\varphi-(\varphi')^2+\varphi^3\psi}{\varphi}.
\end{equation}
\end{enumerate}
\end{lem}

\noindent{\bf Proof of Proposition \ref{prop:F2andtau}.}
Consider the Toda equations (\ref{Toda:sigma}) and (\ref{Toda:tau}). Let us put
\begin{equation}
\tilde{\tau}_n = \frac{\sigma_n}{\sigma_{-1}}=\frac{\tau_n}{\tau_{-1}}
\end{equation}
so that $\tilde\tau_{-1}=1$. Then it is easy to derive the Toda
equation for $\tilde\tau_n$:
\begin{equation}
\tilde\tau_n''\tilde\tau_n - (\tilde\tau_n')^2=
\tilde\tau_{n+1}\tilde\tau_{n-1}-\frac{\psi''\psi-(\psi')^2+\varphi\psi^3}{\psi^2}\tilde\tau_n^2,
\end{equation}
\begin{equation}
 \tilde\tau_{-2}=\frac{\psi''\psi-(\psi')^2+\varphi\psi^3}{\psi},\quad\tilde\tau_{-1}=1,
\quad\tilde\tau_0=\frac{1}{\psi}.
\end{equation}
We have the determinant formula for $\tilde\tau_n$ as,
\begin{equation}
 \tilde\tau_n=\left\{
\begin{array}{ll}
\det(\tilde{a}_{i+j-2})_{i,j\leq n+1}&n>0,\\
1, &n=0,\\
\det(\tilde{b}_{i+j-2})_{i,j\leq |n|-1}&n<0,
\end{array}
\right.\quad,
\end{equation}
\begin{eqnarray}
\tilde{a}_n&=&\tilde{a}_{n-1}^\prime + \frac{\psi''\psi-(\psi')^2+\varphi\psi^3}{\psi}
\sum_{i+j=n-2\atop i,j\geq 0}\tilde{a}_i\tilde{a}_j,
\quad \tilde{a}_0=\frac{1}{\psi},\\ 
\tilde{b}_n&=&\tilde{b}_{n-1}^\prime + 
\frac{1}{\psi}\sum_{i+j=n-2\atop i,j\geq 0}\tilde{b}_i\tilde{b}_j,
\quad \tilde{b}_0=\frac{\psi''\psi-(\psi')^2+\varphi\psi^3}{\psi}.
\end{eqnarray}
Now it is obvious from Lemma \ref{lem:recdn} that
\begin{equation}
d_j=\tilde{b}_{j-2}\quad (j\geq 2),\quad\kappa_n=\tilde\tau_{n}\quad (n< 0),
\end{equation}
which proves equation (\ref{kappaandtau}). Equation (\ref{thetaandtau}) can be proved in similar manner.\qed

\vskip 0.1 cm
We remark that the mysterious relations among the $\tau$
functions and the solutions of isomonodromic problem in Proposition
\ref{prop:FandG} and \ref{prop:F2andtau} should eventually originate
from the symmetry of P$_{\rm II}$, but their meaning is not sufficiently
understood yet.

\section{Summability of the generating function}

To study the growth as $n\to\infty$ of the coefficients $a_n(x)$ (or $b_n(x)$) in (\ref{coefs}) we use a
theorem proved in \cite{HS}, based on a result by Ramis \cite{ram}.

\begin{thm}\label{hs}
Consider the following nonlinear differential equation in the variable $s$
\begin{equation}\label{Heq}
s^{k+1} \frac{{\rm d}H}{{\rm d} s}= c(s) H + s\, b(s,H),
\end{equation}
where $k$ is a positive integer, $c(s)$ is holomorphic in the neighbourhood of $s=0$ and $c(0)\neq 0$, and $b(s,H)$ is holomorphic in the neighbourhood of $(s,H)=(0,0)$. Then equation (\ref{Heq}) admits one and only one formal solution $H_f(s)$ of the form $H_f(s)=\sum_{n=1}^\infty a_n s^n$. Moreover 
$H_f$ is $k$-summable in any direction ${\rm arg}(s)=\vartheta$ except a finite number of values $\vartheta$. Furthermore the sum of $H_f(s)$ in the direction ${\rm arg}(s)=\vartheta$ is a solution of equation (\ref{Heq}).
\end{thm}

Equation (\ref{riccati:Ft}) can be put into the form (\ref{Heq}) by changing the variable $t=\frac{1}{s}$ and taking $H=F-a_0$. We obtain equation (\ref{Heq}) with $k=3$ and
$$
c(s) = \frac{1}{2}\left(1-\frac{\psi''}{\psi} s^2 -2 s^3\right),
$$
$$ 
b(s,H) = -\frac{1}{2}\left(\varphi\,(\psi''/\psi \, s+2 s^2) +\varphi'+ s(\psi-\psi'\, s)\varphi^2+ 
2 s\, \varphi(\psi-\psi' \, s)H+ s(\psi-\psi'\,s)H^2.\right)
$$
Applying theorem \ref{hs}, we obtain that equation (\ref{riccati:Ft}) admits one and only one formal solution $F_\infty(t)$ of the form $F_\infty(t)=\sum_{n=0}^\infty a_n t^{-n}$. This formal solution 
is $3$-summable in any direction ${\rm arg}(t)=\vartheta$ except a finite number of values $\vartheta$ and its sum in the direction ${\rm arg}(s)=\vartheta$ is a solution of equation (\ref{riccati:Ft}). The definition of $k$-summability implies that $F_\infty(t)$ is Gevrey of order $3$, namely, for each $x$ there exist positive numbers $C(x)$ and $K(x)$ such that
$$
|a_n(x)|< C(x) (n!)^{1/3} K(x)^n,\qquad \hbox{for all}\, n\geq 1.
$$
Clearly, one can prove a similar result for the coefficients $d_n$ of the second
formal solution $F^{(2)}$ of equation (\ref{riccati:Ft}). One has to apply
theorem  \ref{hs} to a new series $H$ defined as $H(s)=s^2 F^{(2)}-d_0$.


\begin{thebibliography}{99}
\bibitem{AS}
M.~J. Ablowitz and H. Segur.
\newblock Exact linearization of a Painlev\'e transcendent.
\newblock {\em Phys. Rev. Lett.}, 38:1103--1106, 1977.
%
\bibitem{Ai}
H. Airault.
\newblock Rational solutions of Painlev\'e equations.
\newblock {\em Studies in Applied Mathematics}, 61:31--53, 1979.
%
%\bibitem{CM}
%P.A. Clarkson and E.L. Mansfield.
%\newblock The second Painelv\'e equation, its hierarchy and associated 
%special polynomials.
%\newblock {\em Nonlinearity}, 16:R1--R26, 2003.
%
%
\bibitem{Er}
N.P. Erugin.
\newblock On the second transcendent of Painlev\'e.
\newblock{\em Dokl. Akad. Nauk BSSR}  2:139--142, 1958.
%
\bibitem{FN}
H. Flaschka and A.~C. Newell.
\newblock Monodromy and Spectrum Preserving Deformations I.
\newblock {\em Comm. Math. Phys.}, 76:65--116, 1980.
%
\bibitem{HS}
P.F. Hsieh and Y. Sibuya.
\newblock Basic theory of ordinary differential equations.
\newblock {\em Universitext, Springer, New York,}\/ 1999.
%
\bibitem{IKN} K. Iwasaki, K. Kajiwara and T. Nakamura.
\newblock Generating function associated with the rational solutions of
	the Painlev\'e II equation.
\newblock {\em J. Phys. A: Math. Gen} 35:L207--L211, 2002.
%
\bibitem{JMU} M. Jimbo and T. Miwa, 
\newblock Monodoromy preserving deformation of linear ordinary
	differential equations with rational coefficients.II.
\newblock {\em Physica} 2D:407--448, 1981.
\bibitem{J1}
M. Jimbo. 
\newblock Monodromy problem and the boundary condition for some
	Painlev\'e equations.
\newblock {\em Publ. RIMS} 18:1137--1161, 1982.
\bibitem{J2}
M. Jimbo. 
\newblock{\em Unpublished work.}
%
\bibitem{KO:p2} K. Kajiwara and Y. Ohta.
\newblock Determinant structure of the rational solutions for the
	Painlev\'e II equation.
\newblock {\em J. Math. Phys. }, 37:4693--4704, 1996.
%
\bibitem{KM:p2_generic} K. Kajiwara and T. Masuda.
\newblock A generalization of the determinant formulae for the solutions
	of the Painlev\'e II equation.
\newblock {\em J. Phys. A: Math. Gen. } 32:3763--3778, 1999.
%
\bibitem{KMNOY:det}K. Kajiwara, T. Masuda, M. Noumi, Y. Ohta and Y. Yamada.
\newblock Determinant formulas for the Toda and discrete Toda equations.
\newblock {\em Funkcial. Ekvac.} 44:291--307, 2001.
%
\bibitem{Murata} Y.~Murata.
\newblock Rational solutions of the second and the fourth 
Painlev\'e equations.  
\newblock {\em Funkcial. Ekvac.}, 28:1--32, 1985.
%
\bibitem{Noumi-Okamoto} M. Noumi and K. Okamoto.
\newblock Irreducibility of the second and the fourth Painlev\'e equations.
\newblock {\em Funkcial. Ekvac.}, 40:139--163, 1997.

\bibitem{O1} 
K. Okamoto.
\newblock Private communication.
%
\bibitem{ram}
J.P. Ramis.
\newblock S\'eries divergentes et th\'eories asymptotiques.
\newblock {\em Soc. Math. France, Panoramas Synth\`eses}, 121: 1--74, 1993.

%
\bibitem{Watanabe-Umemura} H. Umemura and H. Watanabe, 
\newblock Solutions of the second and fourth Painlev\'e equations. I.  
\newblock {\em Nagoya Math. J.}, 148:151--198, 1997.
%
\bibitem{Vo}
A.P. Vorob'ev.
\newblock On the rational solutions of the second Painlev\'e equation. 
\newblock {\em  Differencial'nye Uravnenija}  1:79--81,  1965.
%
\bibitem{Ya}
A.I. Yablonskii.
\newblock {\em Vesti Akad. Navuk. BSSR Ser. Fiz. Tkh. Nauk.}\/ 3:30--35, 1959.
%
\bibitem{Wasow}
W. Wasow.
\newblock Asymptotic expansions for ordinary differential equations.
\newblock {\em Pure and Applied Mathematics, John Wiley \and Sons,
Inc.}, XIV, 1965.


\end{thebibliography}
\end{document}